\documentstyle[aps,epsf]{revtex}

\begin{document}
\title{Radiation Pressure as a Source of Decoherence}
\author{Paulo A. Maia Neto $^{1}$ \thanks{e-mail: \tt pamn@if.ufrj.br}
and Diego A.\ R.\ Dalvit $^{2}$  \thanks{e-mail: \tt dalvit@lanl.gov}}

\address{$^1$ \it Instituto de F\'{\i}sica, UFRJ, Caixa Postal 68528, 
21945-970 Rio de Janeiro, Brazil \\
$^2$ \it T-6, Theoretical Division, MS B288, Los Alamos National
Laboratory, Los Alamos, NM 87545}
\date{\today }
\maketitle

\begin{abstract}
We consider the interaction of an harmonic oscillator with the quantum
field via radiation pressure. We show that a `Schrodinger cat' state
decoheres in a time scale that depends on the degree of `classicality' of
the state components, and which may be much shorter than the relaxation time
scale associated to the dynamical Casimir effect. We also show that decoherence
is a consequence of the entanglement between the quantum states of the
oscillator and field two-photon states. 
With the help of the fluctuation-dissipation theorem, we derive a relation
between decoherence and damping rates valid for arbitrary values of the 
temperature of the field. 
Coherent states are selected by the
interaction as pointer states.
\end{abstract}

\section{Introduction}

Superposition states have an important role in the formalism of 
quantum mechanics. However, they are in flagrant contradiction with our classical world
when the components correspond to macroscopically distinguishable states.
The reason why these states are not encountered round the corner is
decoherence, a process by which the interaction between the degrees of freedom
of the system in question with any other degrees of freedom, either internal
or external (the so-called environment), leads to a suppression of the 
coherence between the components of the superposition
\cite{physicstoday}. Even if this coupling
is very weak, the decoherence rate may be huge, resulting in a very fast
decay of these ``weird'' states and in the emergence of the classical world.
Recent developments in technology now allow one to study in real-time
the process of decoherence in the lab. For example, over the past
several years techniques have been developed to generate mesoscopic 
superpositions of motional states of trapped ions \cite{monroe}, and of
photon states in cavity quantum electrodynamics \cite{haroche}. In these cases
decoherence due to the coupling with the ambient reservoirs was observed, 
confirming the expectation that the decoherence rate is faster, 
the larger and more separated the state components are~\cite{luiz}. 
Recently another experiment has succeeded
in ``engineering'' the environment in the context of trapped ions,
studying scaling laws of decoherence theory for a variety of reservoirs in a
wide range of parameters \cite{wineland}.

Usually, decoherence is analyzed in the framework of heuristic models that describe 
phenomenological dissipation (the reservoir is taken to be a collection of
harmonic oscillators, coupled linearly to the position operator of the 
system~\cite{caldeira}\cite{all}).
In this paper, we consider instead an ab initio model for decoherence of
a particle  in a harmonic potential, scattering the radiation field 
(at temperature $T$), which then plays the role of the reservoir. 
Starting from first principles, we show that the resulting radiation pressure
coupling with the field transforms an initial pure  superposition state 
of the particle into a statistical mixture. 

Of special relevance is the limit $T=0.$ In this case, the reservoir is the quantum 
vacuum field, which dissipates the mechanical energy of the oscillating 
particle (or `mirror').
This effect is associated to the emission of pairs of photons, the 
so-called dynamical Casimir effect.
Much work has been done on quantum radiation from moving 
mirrors~\cite{review}. Important properties like  
the spectrum of the emitted radiation~\cite{spectrum}, 
the time evolution of the energy-momentum
tensor~\cite{diego-Casimir},
the total radiated intensity and the
dissipative radiation pressure on the particle
(radiation reaction force corresponding to the 
photon emission effect)~\cite{ford}\cite{force-Casimir}
have been considered. 
Here we 
focus our attention on the particle as the system of interest, and show that
decoherence is a consequence of the entanglement between 
particle and field two-photon states.
This result has  fundamental implications, for it shows that
any particle not transparent to the radiation is unavoidably
under the action of decoherence through the radiation 
pressure coupling with vacuum fluctuations. 

The zero temperature limit was  briefly discussed in our previous letter~\cite{prl}. 
This article presents results for finite values of temperature, as well as 
a detailed discussion of the case $T=0.$ The formalism relies on the 1D scalar model
for the field, but extensions to 3D electromagnetic field are also discussed, allowing 
us to address the question of orders of magnitude.
The paper is organized as follows. In section II we start from the 
Hamiltonian model for the radiation pressure coupling, and then 
derive a master 
equation for the particle. In section III we discuss how the
environment selects a prefered basis in the particle's Hilbert space, the 
pointer basis. In Section IV we derive
a general relation between 
 decoherence and damping
rates at arbitrary temperature by means of the fluctuation-dissipation
theorem. The zero  and high temperature 
limits, including extensions to the 3D electromagnetic 
model, are discussed in Sections V and VI. 
Section VII contains our conclusions. Finally,
 in the
appendix an alternative, simpler derivation of the decoherence rate is given, 
which is based on the entanglement between the particle and two-photon states.

\section{The model}

Most treatments of the dynamical Casimir effect are based on the 
assumption 
that the mirror follows a prescribed trajectory, thus 
neglecting the recoil effect. However, in this paper we want to focus
on the mirror as a dynamical quantum system, hence the need to
tackle the full mirror-plus-field dynamics. This has already been addressed
in the framework of 
 linear response theory in order to calculate the 
fluctuations of the position of a dispersive mirror driven by vacuum radiation
pressure~\cite{jr}, and related calculations have been performed in 
Refs.~\cite{jr2,barton,barton2}
to derive mass corrections  caused by the
interaction with the field.

We consider a nonrelativistic partially reflecting mirror of mass
$M$ (with position $q$ and momentum $p$) in a harmonic potential of 
frequency $\omega_0$, and under the action of radiation pressure. We take
a scalar field in 1+1 dimensions, 
which mimics the electromagnetic field modes that propagate along
the direction perpendicular to the plane of the mirror. 
Extensions to the real 3+1 case are analyzed in Secs. V and VI.
We neglect third and higher order
terms in $v/c$, where $v$ is the mirror's velocity (we set $c=1$ hereafter, except when an explicit evaluation of orders of magnitude is required).
Our starting point is the Hamiltonian formalism developed in 
Refs.~\cite{barton} and \cite{barton2} 
(although these references consider a free mirror, the extension to
the harmonic oscillator is straightforward). 
The total Hamiltonian is 
\begin{equation}
H = H_M + H_F + H_{{\rm int}},  \label{H}
\end{equation}
where 
\begin{equation}
H_M = {\frac{p^2}{2M}} + {\frac{M\omega_0^2}{2}} q^2,  \label{HM}
\end{equation}
is the harmonic oscillator Hamiltonian for the mirror, and 
\begin{equation}
H_F = \int {\frac{dx}{2}} \left[\Pi^2 + (\partial_x\phi)^2%
\right] + \Omega \phi^2(x=0)  \label{HF}
\end{equation}
is  the free Hamiltonian for the field $\phi$ and its momentum canonically
conjugated $\Pi=\partial_t \phi$. The second term in the r.-h.-s. of Eq.~(%
\ref{HF}) is associated to the boundary condition of a partially-reflecting
mirror at rest at $x=0.$ In the context of the  plasma sheet
model of Ref.~\cite{barton}, it corresponds to the kinetic energy of
the plasma charged particles. The coupling constant $\Omega$ plays the role
of a transparency frequency, since from Eq.~(\ref{HF}) one derives the
boundary condition 
\[
\partial_x \phi(0^+)-\partial_x \phi(0^-)= 2\Omega \phi(0) 
\]
($\phi$ is continuous at $x=0$), which yields a frequency-dependent
reflection amplitude~\cite{jr}~\cite{barton}: 
\begin{equation}
R(\omega)= -i {\frac{\Omega}{\omega + i\Omega}}. \label{R}
\end{equation}
Finally, the interaction Hamiltonian is given by 
\begin{equation}
H_{{\rm int}} = -{\frac{p {\cal P}}{M}} + {\frac{{\cal P}^2}{2 M}} -{\frac{1%
}{2}} \Omega \phi^2(x=0) {\frac{p^2}{M^2}},  \label{Hint}
\end{equation}
where ${\cal P}= -\int dx \partial_x\phi\,\partial_t \phi$ is the field
momentum operator. $H_{{\rm int}} $ describes, to second order in $v/c,$
  the
modification of the boundary condition for the field due to the motion of
the mirror, which in its turn is affected by the field radiation pressure.
Thus, it provides a coupling between the harmonic oscillator and the field, 
to be treated within perturbation theory. 
The small perturbation parameter  is $v/c,$ and not the transparency frequency
$\Omega,$ which may be arbitrarily large.
The first term in Eq.~(\ref{Hint}) is responsible for the effect of
decoherence to be discussed here. It also accounts for the effects of
emission of photons, dissipation of the mirror's energy, and part of the
mass correction.

We calculate the density matrix $\tilde \rho(t)$ of the combined
mirror-plus-field system using second order perturbation theory, and trace
over the field operators to derive the master equation for the mirror's
density matrix $\rho(t)$ \cite{walls}. We assume that at $t=0$ the mirror and field are
not correlated: $\tilde \rho(0) = \rho(0) \otimes \rho_F,$ where $\rho_F$ is
the density matrix of the field (assumed to be in some steady state; later in this
section we take a thermal equilibrium state). We
find 
\begin{equation}
i \hbar\dot \rho(t) = [H_M,\rho(t)] - \Omega {\frac{\langle\phi^2(0) \rangle 
}{2 M^2}} [p^2, \rho(t)]  \label{master1}
\end{equation}
\[
-{\frac{i }{2\hbar M^2}} \int_0^t dt^{\prime}\Bigl( [ p , [
p^I(-t^{\prime}),\rho(t)]] \sigma(t^{\prime}) + [p,
\{p^I(-t^{\prime}),\rho(t)\}] \xi(t^{\prime})\Bigr), 
\]
where the superscript $I$ indicates the operators to be taken in the
interaction picture. The second term in the r.-h.-s. of Eq.~(\ref{master1})
is the contribution in first-order of perturbation theory of the $p^2$ term
in the interaction Hamiltonian (see Eq.~(\ref{Hint})). It corresponds to a
(cut--off dependent) mass correction given by 
\begin{equation}
\Delta M_1 =\Omega \langle \phi^2(0) \rangle,  \label{M1}
\end{equation}
as already found in Refs.~\cite{jr} and \cite{barton}. The (anti-)symmetric
second order correlation function ($\xi$) $\sigma$ is defined as 
\begin{equation}
\sigma(t)= C(t)+C(-t),  \label{sigma}
\end{equation}
\begin{equation}
\xi(t)=C(t)-C(-t),  \label{xi}
\end{equation}
with 
\begin{equation}
C(t)=\langle {\cal P}^I(t)\,{\cal P}^I(0)\rangle-\langle{\cal P} \rangle^2.
\end{equation}
When computing the correlation functions, we take the unperturbed field, which
corresponds to the static boundary condition (eigenfunctions of $H_F$).

Replacing the free evolution for $p^I(-t^{\prime})$ in (\ref{master1}) yields 
\begin{equation}
i \hbar \dot\rho=[H_M - \frac{\Delta M(t)}{M} \frac{p^2}{2 M},\rho] -
\Gamma(t) [p,\{q,\rho\}] - \frac{i}{\hbar} D_1(t) [p,[p,\rho]] - \frac{i}{%
\hbar} D_2(t) [p,[q,\rho]].  \label{master2}
\end{equation}
The total mass correction is $\Delta M = \Delta M_1 +\Delta M_2,$ where $%
\Delta M_2,$ as well as the remaining coefficients in (\ref{master2}),
originate from the first term in the r.-h.-s. of Eq.~(\ref{Hint}), taken in
second-order perturbation theory. Their meanings are best understood when
writing the Fokker-Planck equation for the Wigner function $W(x,p,t):$ 
\begin{equation}
\partial_t W = -(1- \Delta M/M){\frac{p}{M}}\partial_x W +M\omega_0^2 x
\partial_p W +2 \Gamma \partial_x(xW)+ D_1 {\frac{\partial^2}{\partial x^2}}%
W - D_2 {\frac{\partial^2}{\partial x\partial p}}W.  \label{fokker}
\end{equation}
$\Delta M_2$ and the damping coefficient $\Gamma$ are calculated from the
anti-symmetric correlation function: 
\begin{equation}
\Delta M_2(t) = {\frac{i}{\hbar}} \int_0^t dt^{\prime}\cos(\omega_0
t^{\prime})\,\xi(t^{\prime}),  \label{M2}
\end{equation}
\begin{equation}
\Gamma(t) = {\frac{i \omega_0}{2 M \hbar}} \int_0^t  dt^{\prime}\sin(\omega_0
t^{\prime})\, \xi(t^{\prime});  \label{Gamma}
\end{equation}
whereas the diffusion coefficients are associated to the symmetric
correlation function: 
\begin{equation}
D_1(t) = {\frac{1}{2 M^2}} \int_0^t dt^{\prime}\cos(\omega_0 t^{\prime})\,
\sigma(t^{\prime}),  \label{D1}
\end{equation}
\begin{equation}
D_2(t) = {\frac{\omega_0}{2 M}} \int_0^t  dt^{\prime} \sin(\omega_0 t^{\prime})\,
\sigma(t^{\prime}).  \label{D2}
\end{equation}

We assume the field to be in a thermal state (temperature $T$), and
take the following strategy to calculate the momentum correlation
functions.
The time derivative of the field momentum is minus the radiation pressure
force on the mirror~\cite{barton}: 
\begin{equation}
{\frac{d{\cal P}^{I}}{dt}}=2\Omega \phi (0,t){\bar{\partial}_{x}\phi(0,t)},
\label{force}
\end{equation}
where ${\bar{\partial}_{x}\phi(0,t)}=[\partial _{x}\phi (0^{+})+\partial
_{x}\phi (0^{-})]/2.$ Using Eq.~(\ref{force}), we calculate $C(t)$ by
integrating the correlation function of the field calculated at $x=0:$ 
\begin{eqnarray}
C(t) &=& 
(2\Omega )^{2}\int_{-\infty }^{t}dt_{1}\int_{-\infty }^{0}dt_{2}
\left[
\langle \phi (0,t_{1}) {\bar{\partial}_{x}\phi(0,t_1)} 
        \phi (0,t_{2}) {\bar{\partial}_{x}\phi(0,t_2)} \rangle  - 
\right. \nonumber \\
&& ~~~~~~~~~
\left.
\langle \phi (0,t_{1}) {\bar{\partial}_{x}\phi(0,t_1)} \rangle 
\langle \phi (0,t_{2}) {\bar{\partial}_{x}\phi(0,t_2)} \rangle 
\right]
\label{der}
\end{eqnarray}
The equal-time second-order correlation function in~(\ref{der})
corresponds to the force on the static (single) mirror. 
It vanishes since the vacuum radiation pressures 
exerted on each side of the mirror are in perfect equilibrium. 
On the other hand, the fourth-order correlation function 
may be expressed as a sum of second order correlation functions (with the fields 
taken at different times), which 
are calculated with the help of the normal mode expansion for the field.
They are directly 
connected to the average number of photons $n_{\omega }=1/[\exp
(\hbar \omega /T)-1]$ in the mode of frequency $\omega$ 
at temperature $T$ (we take the Boltzmann constant $%
k_{\rm B}=1$). It is useful to write the result in the Fourier domain, the
Fourier transform of the anti-symmetric correlation function $\xi (t)$ being
defined as 
\begin{equation}
\xi \lbrack \omega ]=\int dt\exp (i\omega t)\xi (t).  \label{fourier}
\end{equation}
Eqs.~(\ref{xi}) and (\ref{der}) yield 
\begin{equation}
\xi \lbrack \omega ]=\xi ^{0}[\omega ]+\xi ^{T}[\omega ],\label{xitot}
\end{equation}
where 
\begin{equation}
\xi ^{0}[\omega ]=(2/\pi )\hbar^2\Omega \zeta(\omega/\Omega),
\label{cvacres}
\end{equation}
with 
\begin{equation}
\zeta (u)=\ln (1+u^{2})/(2u)+(\arctan u)/u^{2}-1/u,\label{zetadef}
\end{equation}
represents the correlation function at $T=0$ (vacuum fluctuations),
whereas 
\begin{equation}
\xi ^{T}[\omega ]={\frac{2\hbar^2\Omega ^{2}}{\pi \omega ^{2}}}\int_{0}^{\infty
}d\omega ^{\prime }{\frac{\omega ^{\prime }}{\omega ^{\prime }{}^{2}+\Omega
^{2}}}\left[ G(\omega ,\omega ^{\prime })-G(-\omega ,\omega ^{\prime })%
\right] ,  \label{xiT}
\end{equation}
with 
\begin{equation}
G(\omega ,\omega ^{\prime })=|\omega ^{\prime }-\omega |\left( n_{|\omega
^{\prime }-\omega |}-\epsilon (\omega ^{\prime }-\omega )n_{\omega ^{\prime
}}\right) ,\label{G}
\end{equation}
represents the thermal fluctuations ($\epsilon $ is the sign function).

Symmetric and anti-symmetric correlation functions for a system in thermal
equilibrium are related in a very general way~\cite{kubo}\cite{JRT}: 
\begin{equation}
\sigma[\omega] = {\frac{ \xi[\omega]}{\tanh({\frac{\hbar \omega}{2 T}})}}.
\label{fd1}
\end{equation}
According to Eqs.~(\ref{Gamma}) and (\ref{D1}), this result provides a
general relation between diffusion and damping, in the spirit of the
fluctuation-dissipation theorem. This relation is particularly simple for the
asymptotic values of the coefficients $\Gamma (t)$
and $D_{1}(t)$ at $t\rightarrow \infty.$
Since the  integrands in Eqs.~(\ref{Gamma}) and (%
\ref{D1}) are even functions of time, we may
extend the integration range to $-\infty ,$ yielding 
\begin{equation}
\Gamma ={\frac{\omega _{0}}{4M\hbar }}\xi \lbrack \omega _{0}]
\label{Gammaxi}
\end{equation}
and 
\begin{equation}
D_{1}={\frac{1}{4M^{2}}}\sigma \lbrack \omega _{0}].  \label{D1sigma}
\end{equation}
Thus, the asymptotic values of $\Gamma $ and $D_{1}$ are
directly connected to the fluctuations at the mechanical frequency $\omega
_{0},$ allowing us to 
derive, from Eq.~(\ref{fd1}), a simple and general relation between 
these two coefficients.
On the other hand, no such simple connection exists for the remaining
time-dependent coefficients, $\Delta M_{2}$ and $D_{2},$ whose asymptotic
values result from the joint contribution of the whole spectrum of
fluctuations~\cite{prl}.

Combining Eqs.~(\ref{fd1})--(\ref{D1sigma}), we find
\begin{equation}
D_1 = {\frac{\hbar}{M\omega_0}} {\frac{\Gamma}{\tanh({\frac{\hbar \omega_0}{2 T}})%
}} ,  \label{fd2}
\end{equation}
a clear manifestation of the fluctuation-dissipation theorem. 
According to Eq.~(\ref{fd2}), the 
temperature dependence of the diffusion coefficient is determined, 
apart from the $T$ dependence of the 
damping coefficient $\Gamma$ (to be discussed later),
by the relative importance of the thermal fluctuations
(and their corresponding energy $T$) with respect to quantum fluctuations
(and their corresponding zero-point energy $\hbar \omega_0/2$). In the
high-temperature limit, $T\gg \hbar \omega_0/2,$ Eq.~(\ref{fd2}) yields $D_1 = 2 T \Gamma
/(M \omega_0^2).$
In the theory of Brownian motion, $\Gamma$ is a $T$ independent phenomenological
constant, and hence the diffusion coefficient is a linear function of
temperature in this limit. Here, however, $\Gamma$ has an explicit temperature dependence, 
to be analyzed in Sec.~V.

From Eq.~(\ref{fd2}), we shall derive a relation between decoherence and damping
time scales, valid for any temperature $T.$ Before considering a specific superposition 
state, however, we discuss, in the next section, the degree of sensitivity of different states
in the Hilbert space to the action of decoherence. We also analyze in more detail the 
precise meaning of $t \rightarrow\infty$ (in the
particular case of $T=0$), in order to know how fast the coefficients approach their 
asymptotic values. From Eqs.~(\ref{Gamma}) and (\ref{D1}) alone it may be shown, in a 
general way, that
a {\it sufficient} condition is $t\gg 1/\omega_0,$ but in some cases the convergence may
be much faster.


\section{Pointer states}

Different criteria have been introduced in the literature in order to
find out the states in the Hilbert space that are most robust under the
interaction with the environment and behave more classically
\cite{zurek,Paz,wiseman}.
Here we shall follow the one introduced by Zurek, the so-called 
``predictability sieve''. The idea is to take every possible state of the
Hilbert space, calculate its entropy, and order the states in a tower 
according to increasing entropy. The most classical states are those that lie 
at the bottom of that tower, and correspond to the most predictable ones.
For these `pointer' states, information loss due to the interaction 
with the environment is minimal. This philosophy is put in quantitative terms
by minimizing the linear entropy  
of the system,
\begin{equation} 
s[\rho] \equiv 1-{\rm tr}\rho^2,
\end{equation}
which is zero for a pure state and greater than zero for a statistical mixture.
In general this is a difficult problem because complicated entanglement
between system and environment develops on account of their mutual 
interaction. So far, results have been successfully derived
assuming that the initial state of the system is pure. Here we follow the same
approach, and
calculate the rate of entropy increase starting from 
the master equation~(\ref{master2}). We assume that the state is nearly pure
at time $t$ to find
\begin{equation}
{\dot s}(t) = 2 \Gamma(t) (s(t)-1)  +
\frac{4 D_1(t)}{\hbar^2} \Delta p^2 + 
\frac{2 D_2(t)}{\hbar^2} \sigma_{q,p}
\label{s1}
\end{equation}
where $\Delta p^2 \equiv \langle p^2 \rangle - \langle p \rangle^2$ 
is the momentum dispersion and 
$\sigma_{q,p} \equiv \langle \{q,p\} \rangle -2 \langle p \rangle 
\langle q \rangle$. Here $\langle \ldots \rangle = {\rm tr}( \ldots \rho)$,
and all operators are evaluated at the same time $t$. 
The first term in (\ref{s1}) leads to a decrease of entropy
$s(t) = 1 - \exp(2 \int_0^t \Gamma(t') dt')$, hence damping tries to localize
the state competing against diffusion. This decrease is independent of the
initial state of the system, and therefore is irrelevant for determining the
pointer states.

We assume that the typical decoherence time scale is much larger than 
the period of free oscillation $2\pi/\omega_0,$ so that we may integrate 
Eq.~(\ref{s1}) to find the entropy at an intermediate time $\tau = 
n\, 2 \pi/\omega_0.$ We take $n \gg 1,$  allowing us to replace the time
dependent coefficients by their constant asymptotic values, but  
assume that $\tau$ is much shorter than the decoherence time scale,
in order to be consistent with the small-entropy approximation underlying 
Eq.~(\ref{s1}).  Moreover, in this weak coupling limit, we
may take the free evolution (corresponding to the harmonic oscillator
Hamiltonian $H_M$) for the mirror's operators  in Eq.~(\ref{s1}).
The correlation function $\sigma_{q,p}$ oscillates around zero,  and then 
does not contribute to $s(\tau),$ whereas the free evolution of 
$\Delta p^2(t)$ 
mixes up position and momentum fluctuations, yielding 
\begin{equation}
s(\tau)= 2 \tau {D_1\over \hbar^2} \left[ (\Delta p)_0^2 +
(M\omega_0)^2 (\Delta q)_0^2- M \hbar \omega_0\right], \label{deltas}
\end{equation}
where 
$(\Delta p)_0^2$ and $(\Delta q)_0^2$ represent the dispersions 
for the initial state. From Eq.~(\ref{deltas}), 
we find that the minimum entropy given the constraint
$\Delta q \Delta p \ge \hbar/2$
is for $\Delta q{}^2=\hbar/(2M\omega_0),$ $\Delta p{}^2=M\hbar\omega_0/2.$
Hence, as in the Caldeira-Leggett model, and for any temperature of the field,
the pointer basis consists of coherent states~\cite{Paz}.
In this weak coupling approximation, 
the minimum value corresponds to
$s(\tau)=0,$ hence the increase of entropy of 
a coherent state is a higher order effect. 

\begin{figure}
\centering \leavevmode
\epsfxsize=7cm
\epsfbox{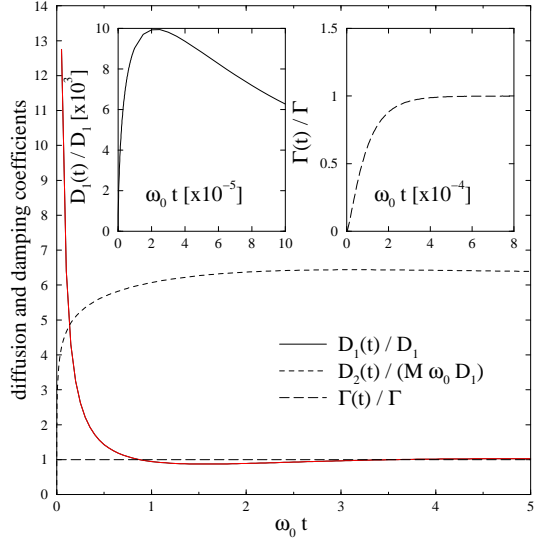}
\caption{\small{Diffusion and damping coefficients for zero temperature  
as  functions of time
in the perfectly-reflecting limit,  $\omega_0/\Omega=10^{-4} \ll 1$. Here 
$D_1= \hbar^2 \omega_0/ 12 \pi M^2$ 
and $\Gamma=\hbar \omega_0^2/12 \pi M$
are the asymptotic limits of $D_1(t)$ and $\Gamma(t)$. The insets
show the behavior of these two time-dependent coefficients for short
times.}}
\end{figure}

The crucial approximation in the derivation  of~(\ref{deltas}) from
(\ref{s1}) is the replacement  of the time dependent coefficients 
by their finite, constant asymptotic values. It is instructive 
to analyze in more detail the behavior of the coefficients and 
its connection with entropy production. As an example, we take 
$T=0,$ and consider first 
the perfectly-reflecting limit,
which corresponds to
$\omega_0\ll \Omega,$ 
for in this case the relevant field modes have frequencies much smaller than the 
mirror's transparency frequency. 
In Fig.~1 we plot the diffusion and damping
coefficients as functions of $\omega_0 t$ for $\omega_0/\Omega=10^{-4}$
and $T=0.$ The damping coefficient $\Gamma$ 
approaches its asymptotic value very fast, 
for $t\sim 1/\Omega,$ 
whereas
$D_1(t)$ develops an initial jolt for times
of the order of $\Omega^{-1}$ and then decreases to the asymptotic value
$(D_1)_{\rm perf}=\hbar^2 \omega_0/(12 \pi M^2)$ 
for $t \sim 1/\omega_0.$
When we integrate Eq.~(\ref{s1}) over many 
periods of oscillation, 
the contribution to the entropy of the initial jolt 
is negligible, allowing us to 
replace $D_1$ by its asymptotic
value. 

\begin{figure}[h]
\centering \leavevmode
\epsfxsize=7cm
\epsfbox{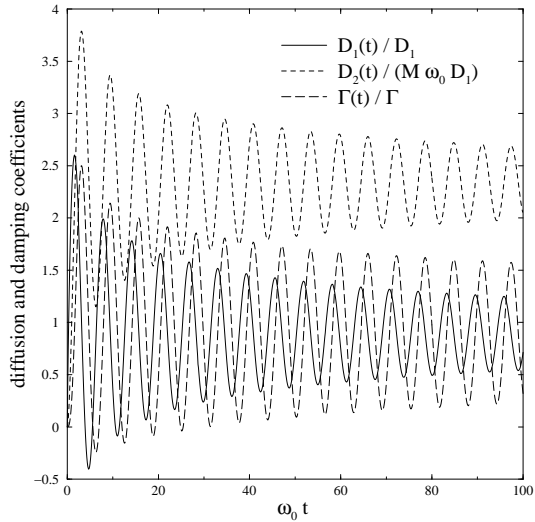}
\caption{\small{Diffusion and damping coefficients for zero temperature
$T=0$ as a function of time
in the high-transmission limit $\omega_0/\Omega=10^{4} \gg 1$. Here
$D_1=\hbar^2 \Omega^2 \ln(\omega_0/\Omega) / 2 \pi M^2 \omega_0$ and
$\Gamma=\hbar \Omega^2 \ln(\omega_0/\Omega)/2 \pi M$ are the asymptotic 
limits of $D_1(t)$ and $\Gamma(t)$.}}
\end{figure}

In Ref.~\cite{energy}, it was shown that no net entropy is produced 
for the Caldeira--Leggett model with an adiabatic environment,
since all the time-dependent coefficients are
oscillating functions around a zero mean. At first sight, 
the same would happen in our model when 
 $\omega_0 \gg \Omega,$ for in this case the dominant field frequencies 
 are slow with respect to the mirror's translational time scale. 
However,  as discussed in Section V,
 the spectral density $\xi^0(\omega)$ decays 
too slowly for $\omega \gg \Omega,$ and as a consequence
field frequencies of the order of $\omega_0$ provide a significant
contribution even in this limit. Thus, one cannot ascribe a
frequency cut-off to the environment such that the typical frequency of the
system $\omega_0$ is much greater than the maximum frequency of the 
environment.
Therefore, the vacuum field does not behave adiabatically in the 
sense of~\cite{energy}.
In our case instead, the diffusion coefficients oscillate around
a non-zero value,  leading to a net entropy increase.
In Fig.~2, we plot the diffusion and damping coefficients as functions of 
$\omega_0 t$ for $\omega_0/\Omega=10^{4}$ and $T=0.$
They oscillate around their asymptotic values with 
(angular) frequency $\omega_0$ and with an  amplitude of
oscillation that decays in a time  $t \sim 1/\Omega\gg 1/\omega_0.$
The oscillatory 
terms do not contribute to the entropy increase when we average over 
many oscillations.
Hence Eq.~(\ref{deltas}) also holds in this case, although 
the rate of entropy increase is much smaller than in the 
perfectly-reflecting limit.


\section{Decoherence versus damping}

In this section, we derive a general relation between damping and
decoherence time scales, starting from the fluctuation-dissipation result 
given by Eq.~(\ref{fd2}).
As an extreme case of decoherent dynamics, 
we consider a superposition of two 
coherent states, since they correspond to the pointer states, according to
the results of Sec.~III.
Specifically, 
we take at $t=0$ the even superposition
state $|\psi \rangle_{\rm e} =(|\alpha \rangle
+|-\alpha \rangle )/\sqrt{2},$  with 
$\alpha =iP_{0}/\sqrt{2M\hbar \omega
_{0}},$ so that
the coherent states are initially along the
momentum axis in phase space, and $\pm P_{0}$ are the average values of momentum of the
components at $t=0.$ We also assume that $|\alpha |\gg 1,$ hence the
average energy of the state components is much larger than the zero-point
energy. The corresponding Wigner function is 
\begin{equation}
W=W_{m}+{\frac{1}{\pi \hbar }}\exp \left[ -{\frac{q^{2}}{2(\Delta q_{0})^{2}}%
}-{\frac{p^{2}}{2(\Delta p_{0})^{2}}}\right] \cos (\frac{2P_{0}q}{\hbar }),
\label{int}
\end{equation}
where $\Delta q_{0}=\sqrt{\hbar /(2M\omega _{0})}$ and $\Delta p_{0}=\hbar
/(2\Delta q_{0})$ are the position and momentum uncertainties of the ground
state. $W_{m}$ corresponds to the statistical mixture 
\begin{equation}
\rho _{m}=(|\alpha \rangle \langle \alpha |+|-\alpha \rangle \langle -\alpha
|)/2. \label{defrho}
\end{equation}
In phase space, $W_{m}$ has two Gaussian peaks along the momentum axis at $%
\pm P_{0}$. $\rho _{m}$ is a classical state in the sense that $W_{m},$
being positive defined, may be interpreted as a probability distribution in
phase space. On the other hand, the nonclassical nature of the superposition
state is featured by the remaining term in Eq.~(\ref{int}), representing the
coherent interference between the two state components, and which oscillates
into negative values along the position axis.

Diffusion along position, associated to the coefficient $D_1,$ averages out
the oscillations of the interference term at a rate $1/t_{\rm d},$ to be
calculated from the Fokker-Planck equation~(\ref{fokker})~\cite{foot1}. According 
to Eq.~(\ref{int}), the oscillations
are faster the higher the value of $P_0,$ so that $t_{\rm d}$ is a decreasing
function of $|\alpha|.$ 
As in Section III, we assume that decoherence is very slow,
$1/t_{\rm d} \ll \omega_0,$ so that several free 
oscillations take place before coherence is lost. 
In this limit, the particle has enough time to probe the harmonic potential before
diffusion takes place, and as a consequence decoherence is governed by the 
asymptotic value of $D_1,$ which is directly connected to
the field flucutuations at the frequency of oscillation $\omega_0,$ 
according to Eq.~(\ref{D1sigma}).
This condition 
holds for most experiments, where mesoscopic superpositions are employed so 
as to render decoherence slow enough to be measured~\cite{haroche}\cite{wineland}. 
Moreover, it 
always applies in the case of vacuum radiation pressure ($T=0$), as shown in Sec.~V.
Diffusion is maximum when the state components are along the momentum axis:
from (\ref{int}), we find $\partial^2_q
W\approx -(2 P_0/\hbar)^2 W;$ and vanishes when the two wavepackets  
reach the turning points in the harmonic potential. The
average over many oscillation yields
\begin{equation}
{\frac{1}{t_{\rm d}}} = -{1\over 2} D_1 {\left(\partial^2_q
W\over W\right)_{\rm max}} = {\frac{ 2 P_0^2 D_1}{\hbar^2}},  \label{final}
\end{equation}
that combined with the fluctuation-dissipation theorem, Eq.~(\ref{fd2}), yields the
following result for the decoherence time $t_{\rm d}:$ 
\begin{equation}
t_{\rm d} = {\frac{1}{4 |\alpha|^2}} \tanh({\frac{\hbar \omega_0}{2 T}}) {\frac{1}{%
\Gamma}}.  \label{res1}
\end{equation}

A $T=0$ (or more generally, for $T\ll \hbar \omega _{0}$), Eq.~(\ref{res1})
yields $t_{d}=1/(4|\alpha |^{2}\Gamma ).$ This result may be written in
terms of the distance $\Delta P=2P_{0}$ between the two components in phase
space at $t=0,$ or in terms of the distance $\Delta Q=\Delta P/(M\omega _{0})$
at $t=\pi /2\omega _{0}:$ 
\begin{equation}
t_{d}=4\left( {\frac{\Delta p_{0}}{\Delta P}}\right) ^{2}{\frac{1}{\Gamma }}%
=4\left( {\frac{\Delta q_{0}}{\Delta Q}}\right) ^{2}{\frac{1}{\Gamma }}.
\label{res2}
\end{equation}
The interpretation of~(\ref{res2}) is clear: decoherence is faster, the more
separated the state components in phase space are. Here the zero-point
fluctuations define the reference of distance in phase space. At high
temperatures, on the other hand, this reference is provided by the thermal de Broglie wavelength $\lambda _{T}=\hbar /\sqrt{2MT}.$ In
fact, (\ref{res1}) yields, for $T\gg \hbar \omega _{0},$ 
\begin{equation}
t_{d}={\frac{\hbar \omega _{0}}{2T}}{\frac{1}{4|\alpha |^{2}}}{\frac{1}{%
\Gamma }}=2\left( {\frac{\lambda _{T}}{\Delta Q}}\right) ^{2}{\frac{1}{%
\Gamma }}. \label{res3}
\end{equation}
Eq.~(\ref{res3}) also shows that the ratio between decoherence and damping
rates is larger at high temperatures by the factor $2T/(\hbar \omega _{0}).$

When written in terms of distances in phase space, the results above are
also valid for more general superposition states, like $(|0\rangle +
|\alpha\rangle)/\sqrt{2}.$ Moreover, their range of validity is not limited
to the radiation pressure coupling considered here. In fact, Eqs.~(\ref{res2}) 
and (\ref{res3}) are in perfect agreement with the results obtained in the
framework of the Caldeira-Legget phenomenological model for quantum
dissipation~\cite{caldeira}. This is hardly surprising, since they rely on
general properties of the correlation functions associated to the
fluctuation-dissipation theorem.
Eq.~(\ref{res1}), which interpolates the low and high temperature limits, 
is also discussed in Ref.~\cite{caldeira}, in the context of a two-level system.
The temperature dependence for the ratio between decoherence and damping times
has a simple interpretation: at $T>0,$ the time
scale for the relaxation of the populations is shorter than $1/\Gamma$ 
exactly by
the factor $\tanh(\hbar \omega_0/2 T),$
on account of the contribution of absorption and stimulated emission.  
Here this factor originates from the general relation between symmetric and
anti-symmetric correlation functions, Eq.~(\ref{fd1}), which is at the heart of the 
fluctuation-dissipation theorem. 

The peculiarities of the radiation pressure model considered here are
contained in the damping rate $\Gamma.$ Rather than a phenomenological
input parameter, it is computed from first principles, first for $T=0$ in
Sec.~V, and then for $T\gg \hbar \omega _{0}$ in Sec.~VI.

\section{Vacuum field}

At $T=0,$ the spectral density is given by Eqs.~(\ref{cvacres}) and (\ref{zetadef}).
This result is more easily obtained from the following argument (a similar
method, applied for the force correlation function, may be found in Refs.~\cite{pamn93} 
and \cite{barton91}). Since ${\cal P}$ is quadratic in the
field operators, the correlation function $C(t)$ may be calculated from the two-photon matrix
elements of the momentum operator as follows: 
\begin{equation}
C(t)={\frac{1}{2}}\int_{0}^{\infty }d\omega _{1}\int_{0}^{\infty }d\omega
_{2}\langle 0|{\cal P}(t)|\omega _{1},\omega _{2}\rangle \langle \omega
_{1},\omega _{2}|{\cal P}|0\rangle .  \label{cvac}
\end{equation}
We have $\langle 0|{\cal P}(t)|\omega _{1},\omega _{2}\rangle =\exp
[-i(\omega _{1}+\omega _{2})t]\langle 0|{\cal P}(0)|\omega _{1},\omega
_{2}\rangle $ since only the annihilation operators contribute, and hence 
\begin{equation}
C[\omega]=\pi \int_{0}^{\infty }d\omega _{1}\int_{0}^{\infty }d\omega
_{2}|\langle 0|{\cal P}(0)|\omega _{1},\omega _{2}\rangle |^{2}\delta
(\omega -\omega _{1}-\omega _{2}).  \label{cvac2}
\end{equation}
Thus, at $T=0$ the fluctuations at the (positive)
frequency $\omega$ originate from two-photon states $|\omega _{1},\omega
_{2}\rangle $ such that $\omega _{1}+\omega _{2}=\omega.$
In the dynamical Casimir effect, the oscillation at the mechanical frequency
$\omega_0$ gives rise to the emission of pairs of photons of frequencies
$\omega_1$ and $\omega_2,$ such that $\omega_0=\omega_1+\omega_2.$ 
On the other hand, according to~(\ref{Gammaxi}), 
$\Gamma$ originates from the 
fluctuations at frequency $\omega_0,$ and hence 
\begin{equation}
\Gamma = {\frac{\pi}{4}}{\frac{\omega_0}{M\hbar}} \int_0^{\infty}
d\omega_1\int_0^{\infty} d\omega_2 |\langle \omega_1,\omega_2|{\cal P}%
|0\rangle|^2 \delta(\omega_0-\omega_1-\omega_2),\label{Gammaa}
\end{equation}
rendering explicit the connection between damping and the photon emission
effect. 
In the appendix, we 
present an alternative derivation of~(\ref{Gammaa}), 
starting from the two-photon emission probabilities and making use of 
energy conservation.

Eq.~(\ref{cvac2}) 
also shows that
$C[\omega ]$ vanishes for negative frequencies, and as a consequence, $\sigma[%
\omega]=\epsilon(\omega) \xi[\omega]$ in agreement with Eq.~(\ref{fd1}).
Finally, the result of Eq.~(\ref{cvacres}) follows from (\ref{cvac2}) by
using again Eq.~(\ref{force})~\cite{barton}. In Fig.~3     , we plot $\zeta
(\omega /\Omega )$ as a function of its argument. According to Eq.~(\ref
{cvacres}), the transparency frequency $\Omega $ defines a scale for the
behavior of the spectrum of fluctuations in vacuum. For $\omega \ll \Omega ,$
the spectrum is linear: $\zeta (\omega /\Omega )\approx \omega /(6\Omega) ,$
and goes to zero slowly, as $\ln (\omega /\Omega )/(\omega /\Omega ),$ for $%
\omega \gg \Omega,$ due to the high-frequency transparency of the mirror.

\begin{figure}
\centering \leavevmode
\epsfxsize=7cm
\epsfbox{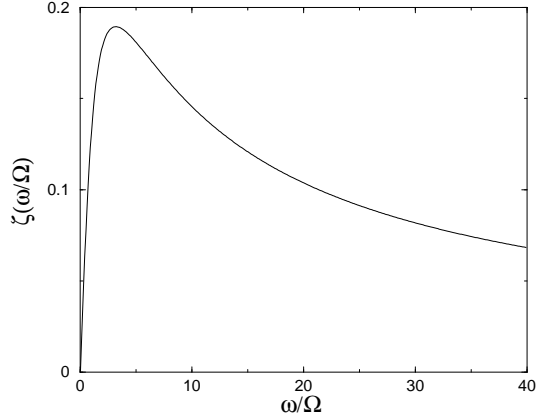}
\caption{\small{Spectral density for  zero temperature.}}
\end{figure}

The damping coefficient at zero temperature is obtained from 
Eqs.~(\ref{cvacres}) and (\ref{Gammaxi}), or alternatively from Eq.~(\ref{Gammaa}): 
\begin{equation}
\Gamma ={\frac{\hbar \Omega \omega _{0}}{2\pi M}}\zeta ({\frac{\omega _{0}}{%
\Omega }}).  \label{Gammares}
\end{equation}
In the perfectly-reflecting limit, $\omega _{0}\ll \Omega ,$
Eq.~(\ref{Gammares}) yields
\begin{equation}
\Gamma ={\hbar \omega _{0} \over 12\pi M}\omega _{0}.\label{Gammaper}
\end{equation}
Thus, the 
damping induced by the Casimir effect is a small perturbation of the 
free harmonic oscillations. The ratio between the zero-point energy and the 
rest energy appearing in (\ref{Gammaper}) is also of the order of the recoil
velocity of the mirror divided by $c,$ which is, as explained in Sec.~II, the 
small parameter of the perturbation approach leading 
to the master equation~(\ref{master2}). For larger
values of $\omega_0/\Omega,$ the damping as given by~(\ref{Gammares}) is still
smaller, since vacuum frequencies of the order of $\omega_0$ are not 
well reflected by the mirror in this case. 

Eq.~(\ref{Gammaper}) is directly connected
to the well-known formula for the dissipative Casimir force on a single
perfect moving mirror~\cite{ford}, $F=\hbar x^{^{\prime \prime \prime
}}/(6\pi ),$ for the equation of motion then reads~\cite{runaway}
\begin{equation}
x^{^{\prime \prime }}=-\omega _{0}^{2}x+{\frac{\hbar x^{^{\prime \prime
\prime }}}{6\pi M}},\label{run}
\end{equation}
whose solution when $\hbar \omega _{0}/M\ll 1$ is 
\[
x=x_{0}e^{-i\omega _{0}t}\exp \left( -\frac{\hbar \omega _{0}^{2}}{12\pi M}%
t\right) .
\]

The decoherence time scale at $T=0$ in the  perfectly-reflecting limit
is derived from Eqs.~(\ref{res2})  and 
(\ref{Gammaper}):
\begin{equation}
t_{\rm d} = {3\over v^2} {2\pi\over \omega_0},\label{res6}
\end{equation}
where $v=P_0/M$ is the initial velocity of the wavepackets. 
Being of the order of $(v/c)^2,$ the decoherence rate is very small
at $T=0$ (or, at any rate, in the nonrelativistic limit 
considered here). Since $\omega_0 t_{\rm d} \gg 1,$ decoherence 
is the cumulative effect of several free oscillations in the 
harmonic well, which justifies the approach employed in the derivation of
(\ref{final}) and the use of the asymptotic value for $D_1(t).$

In order to further understand how the dynamical Casimir effect
engenders decoherence, we present, in the appendix, an alternative
approach, where we follow the evolution of the complete 
oscillator-plus-field quantum state. 
It shows that the superposition state
decoheres because the two wavepacket components oscillating out-of-phase
yield  amplitudes for emission of photon pairs
with opposite signs. As a consequence,  an entangled mirror-plus-field state 
  is developed, given by
\begin{equation}
|\Psi\rangle_{\Delta\! t} = B({\Delta\! t}) |\psi\rangle_{\rm e} 
\otimes |0\rangle + {\frac{1}{2}}
\int_0^{\infty}d\omega_1\int_0^{\infty}d\omega_2\, b(\omega_1,\omega_2;{\Delta\! t})
\,|\psi\rangle_{\rm o} \otimes|\omega_1,\omega_2\rangle, \label{Psit}
\end{equation}
where $|\psi\rangle_{\rm o} = (|\alpha \rangle -|-\alpha \rangle )/\sqrt{2} 
$ is the odd superposition state, $b(\omega_1,\omega_2;{\Delta\! t})$ is the
amplitude for emission of a photon pair with frequencies $\omega_1$ and $\omega_2$
during $\Delta t$ (the explicit 
expressions are given in the appendix),
and $B$ is the amplitude for persistence in the vacuum state: 
\begin{equation}
|B(\Delta\! t)|^2= 1 -{\frac{1}{2}} \int_0^{\infty} d\omega_1
\int_0^{\infty} d\omega_2
|b(\omega_1,\omega_2,\Delta\! t)|^2.  \label{b1b}
\end{equation}

The density operators of the odd and even superposition states differ by the sign
of the interference term, $\rho_{\rm int}=\rho - \rho_m$ ($\rho_m$
is defined in Eq.~(\ref{defrho})).  Accordingly, when computing the 
reduced density matrix of the mirror, $\rho({\Delta\! t}) = {\rm tr}_{\rm field}(|\Psi\rangle_{\Delta\! t}
\langle\Psi|),$ the contribution of the two-photon states in Eq.~(\ref{Psit})
reduces the coherence of the state. With the help of Eq.~(\ref{b1b}), we find
\begin{equation}
\Delta \rho_{\rm int}\equiv \rho_{\rm int}(\Delta\! t)-\rho_{\rm int}(0)=
-{\frac{1}{2}} \int_0^{\infty}d\omega_1\int_0^{\infty}d\omega_2%
\, |b(\omega_1,\omega_2;\Delta \! t)|^2\,
\rho_{\rm int}(0).\label{interf1}
\end{equation}
The two-photon probabilities are proportional to $\Delta t$ and 
connected to the damping rate $\Gamma$ as discussed in the appendix. 
Hence Eq.~(\ref{interf1}) yields
\begin{equation}
{d\rho_{\rm int}\over dt}\approx
{\Delta\rho_{\rm int}\over \Delta\! t}= -{\rho_{\rm int}\over  t_{\rm d}},\label{interf2}
\end{equation}
with $t_{\rm d}$ given by Eq.~(\ref{res2}).

In this derivation, the expression for the emission amplitudes {\it per se} are not
of any relevance  --- only its connection with the damping rate 
$\Gamma$
is important. This connection is based on the principle of energy conservation: 
the energy of the oscillator is damped at the rate at which energy is radiated. 
Since this argument also holds for the real 3+1 
electromagnetic field, we may 
extend our results  by replacing the 3D result for $\Gamma$ into Eq.~(\ref{res2}).
The dissipative dynamical Casimir force on an oscillating (frequency
$\omega_0$) perfectly-reflecting sphere  was obtained 
in Ref.~\cite{pamn93}. Usually, the sphere is very small when compared with the 
wavelength of the relevant vacuum fluctuations, which are of the order of
$2\pi/\omega_0.$ 
When $\omega_0 R \ll 1,$ the force on the 
 sphere of radius $R$ is given by 
\begin{equation}
F = {-\hbar R^6\over 648 \pi} x^{(9)},\label{F3D}
\end{equation}
where $x^{(9)}$ is the ninth time derivative of the position of the sphere.
Following again the method of Eq.~(\ref{run}), we calculate the damping 
rate $\Gamma$ from the equation of motion. We find
\begin{equation}
\Gamma = {1\over 1296 \pi} {\hbar \omega_0^8 R^6 \over M} 
\end{equation}
showing that 
the coupling with the vacuum field is reduced, as compared with
the 1D case, 
 by the (very small) factor $(\omega_0 R)^6.$  
This reduction factor accounts for the inefficient coupling of the 
small particle, which scatters field modes of very long wavelengths.
Using Eq.~(\ref{res2}), we find that the decoherence time increases 
by the same factor:
\begin{equation}
t_{\rm d} = {324\over v^2} {1\over (\omega_0 R)^6}{2\pi\over \omega_0}, 
\end{equation}
hence decoherence through radiation 
pressure is a tiny effect at $T=0.$
At finite temperatures, however, the effect may be 
significant, as discussed in the next section.

\section{High-temperature limit}

In this section, we compute the damping and decoherence rates when $T\gg \hbar \omega_0.$
In this limit, vacuum fluctuations are negligible when compared with 
thermal fluctuations, and the dominant contribution in Eq.~(\ref{xitot}) 
comes from $\xi^T,$ which is given by Eq.~(\ref{xiT}).
When the temperature is also higher than the cut-off energy $\hbar \Omega,$
all relevant frequencies in~(\ref{xiT}), which are smaller or of the order
of $\Omega,$ are much smaller than $T/\hbar.$ Then, we may take the approximation
$n_{\omega'}\approx T/(\hbar \omega'),$ yielding 
\begin{equation}
\xi^T[\omega_0]=2 {\hbar \Omega T\over \omega_0}.\label{xiT1}
\end{equation}
Replacing (\ref{xiT1}) into  
(\ref{Gammaxi}) yields 
\begin{equation}
\Gamma= {\Omega T\over 2 M},\label{resG1}
\end{equation}
in agreement with the result for the viscous radiation
pressure force obtained in Ref.~\cite{JRT}: $F=-\Omega T\dot{q}(t).$

From a practical point-of-view, the opposite limit, $\hbar \omega_0\ll T\ll \hbar\Omega$
is more interesting for particles that scatter visible light
($\Omega \sim 10^{16} {\rm Hz}$). 
In this case, the corresponding reflectivity amplitude $R(\omega)$ 
is approximately constant for the field modes whose 
frequencies are smaller or of the order of $T/\hbar.$ As a
consequence, we may neglect the Lorentzian fall-off in~(\ref{xiT}). 
Moreover, we replace the thermal photon number $n_{\omega'-\omega_0}$ in~(\ref{G}) 
by 
\begin{equation}
n_{\omega'-\omega_0}\approx \left[ \exp(\hbar\omega'/T)(1-\hbar\omega_0/T)-1\right]^{-1}.
\end{equation}
Neglecting second and higher order terms in $\hbar \omega_0/T,$
we find
\begin{equation}
G(\omega_0,\omega')= {\omega' e^{\hbar\omega'\over T}\over (e^{\hbar\omega'\over T}-1)^2} 
{\hbar \omega_0\over T}.\label{Ga}
\end{equation}
From~(\ref{xiT}) and (\ref{Ga}) we find 
\begin{equation}
\xi^T[\omega_0]={4\pi\over 3} {T^2\over \omega_0},\label{xiexp1}
\end{equation}
and then
\begin{equation}
\Gamma= {\pi\over 3} { T^2\over  M\hbar},\label{resG2}
\end{equation}
which is also in agreement with Ref.~\cite{JRT}. It corresponds to the 
high-temperature, perfectly-reflecting limit. Here $T$ plays the role of frequency cut-off
instead of $\Omega,$ so that the resulting damping rate is independent of the latter. 

The dissipative force in the high temperature limit may be interpreted as the effect of
Doppler shift of the reflected thermal photons~\cite{JRT}.  For a 
photon of frequency $\omega,$ the frequency shift is $\Delta\omega=\pm 2\omega {\dot q},$
the plus and minus signs applying for 
counter and co-propagating cases.  
Hence the motion gives rise to an unbalance between the radiation pressure exerted 
on each side
of the mirror, corresponding to a  momentum transfer  $\Delta P = 2 \Delta E\,  {\dot q},$
where $\Delta E$ is the reflected energy during a time interval $\Delta t.$ 
In terms of the 
density of modes $g(\omega),$ we have
\begin{equation}
\Delta E = \int_0^{\infty} d\omega |R(\omega)|^2 g(\omega) n_{\omega}\, \hbar \omega,
\label{deltaE}
\end{equation}
where  $|R(\omega)|^2$ represents the mirror reflectivity (the reflection amplitude
$R$ is given by Eq.~(\ref{R})). From $\Delta E,$ the friction force is obtained through 
\begin{equation}
F=-2{\Delta E\over \Delta t} {\dot q}.\label{fdelta}
\end{equation}

In the 1D case, the density of modes is frequency independent: 
$g(\omega) d \omega= (L/\pi) d\omega$, where $L=\Delta t$ is the length of the
quantization box. 
When replaced into Eq.~(\ref{deltaE}), this result leads, with the help of
(\ref{fdelta}), to  expressions for the force in agreement with our results for
$\Gamma,$ except for a factor of $2$ when $\hbar \omega_0\ll T \ll \hbar \Omega$ \cite{foot2}. 
In the 3D case, on the other hand, we have 
\begin{equation}
g(\omega) d\omega= \frac{V}{\pi^2}  \omega^2 d\omega,\label{g3D}
\end{equation}
where $V=A \Delta t$ is the quantization volume, $A$ being the surface of the 
mirror (in this case, for simplicity, we assume a flat rather than spherical 
mirror). 
In the limit $\hbar \omega_0\ll T \ll \hbar \Omega,$
Eqs.~(\ref{deltaE}) and (\ref{g3D}) yield
\begin{equation}
{\Delta E\over \Delta t} = {\hbar A\over  \pi^2} \int_0^{\infty} 
d\omega {\omega^3\over \exp\left(\hbar \omega/T\right)-1} = {\pi^2\over 15}
{A T^4\over \hbar^3}.\label{stefan}
\end{equation}
As expected,
the reflected power features the $T^4$ dependence of the Stefan-Boltzmann law, 
since it is proportional to the total blackbody radiation energy in this limit.  
The friction force is found by replacing Eq.~(\ref{stefan}) 
into (\ref{fdelta}).
The resulting damping coefficient is given by
\begin{equation}
\Gamma = {\pi^2\over 15} {A\over \hbar^3} {T^4\over M}.\label{gamma3d}
\end{equation}
Since we have neglected diffraction at the borders of the mirror,
this result only applies when the mirror is much larger than the
thermal photon wavelength $\lambda_{\rm th}=2\pi \hbar/T.$  

The decoherence time is then found by replacing~(\ref{gamma3d}) 
into~(\ref{res3}), which connects damping and decoherence in the 
high-temperature limit (we re-introduce
the speed of light $c$ in order to allow an evaluation of the orders of
magnitude):
\begin{equation}
t_{\rm d} = \frac{15}{32 \pi^7}  
{\lambda_{\rm th}^5\over c A  \Delta\! Q^2}.\label{restemp}
\end{equation}
As a
numerical example, we take $T=50 K,$ which gives 
$\lambda_{\rm th} = 2.9 \times 10^{-4} {\rm m},$ and $A=1 {\rm mm^2}.$
In this case, diffraction is negligible, and Eq.~(\ref{restemp}) yields
$t_{\rm d}[{\rm s}]= 1.0 \times 10^{-24}/(\Delta\! Q^2[{\rm m^2}]) ,$ showing that
decoherence is very fast even when the distance between the wavepackets is,
for instance, in the nanometer range --- in this case the decoherence time is
of the order of a micro-second. 
Since $t_{\rm d}$ scales as $1/T^5,$ it is still shorter, by a factor $\approx 8\times 10^3,$ 
at room temperature.  Note, however, 
that
Eq.~(\ref{restemp}) only applies when 
$\omega_0 t_{\rm d} \gg 1,$ the basic assumption that allowed us to 
relate decoherence and damping time scales with the help of the fluctuation-dissipation
theorem~\cite{foot3}.

\section{Conclusions}

As in the Caldeira-Legget model~\cite{Paz}, coherent states are the most
robust when the radiation pressure coupling with the 
quantum field is considered. 
This is amazingly in line with their well-known status
of `quasi-classical' states, i.e., 
the closest possible representation of a classical 
oscillation in a harmonic potential well. 
Superpositions of coherent states decay into a 
mixture at a rate 
proportional to the damping rate and to the squared
distance in phase space. 
The ratio between decoherence and damping rates is a simple 
hyperbolic increasing function of temperature, which interpolates the 
zero and high temperature limits. It originates from the general relation 
between symmetric and anti-symmetric correlation functions, associated to the 
fluctuation-dissipation theorem. Thus, the particular nature of the model
for the coupling with the reservoir seems to be immaterial, as far as 
the connection between damping and decoherence is concerned.
Note, however, that the validity of this result is limited by the 
assumption that decoherence is slow compared to the free oscillations.   
   
We have shown that the radiation pressure exerted by thermal photons is
a very efficient source of decoherence, although the corresponding 
energy damping effect, associated to the Doppler frequency shift of the 
reflected photons, is usually negligible. 
At $T=0,$ the energy damping is associated to the emission of 
photon pairs (dynamical Casimir effect). The dominant contribution comes
from vacuum fluctuations corresponding to wavelengths of the order of
$2\pi c/\omega_0,$ which is usually much greater than 
the size of the oscillator. As a consequence, the radiation pressure coupling is
inefficient, and both damping and decoherence rates become very small. 
It is however remarkable, from a theoretical point-of-view, that 
the mere inclusion of 
an unavoidable, intrinsically quantum effect, is sufficient (in principle)      
to engender decoherence, and by that means restoring, although in a very long
time scale, the classical world. 

We are grateful to A. Calogeracos and 
G. Barton for correspondence, and to J. Dziarmaga,
A. Lambrecht, M.-T. Jaekel and S. Reynaud  for 
discussions. P. A. M. N. thanks CNPq, PRONEX and FAPERJ for partial financial 
support.

\appendix
\section{Entanglement with two-photon states}

In this appendix, we present an alternative, simpler derivation of the
decoherence time scale at $T=0,$ which shows more clearly how the dynamical Casimir
effect modifies the quantum phase of a superposition state and engenders
decoherence.
Instead of tracing over the field, we follow its evolution during 
many periods of free oscillation. 
We first take, at $t=0,$ 
the mirror-plus-field state $|\alpha\rangle\otimes |0\rangle$ 
($|0\rangle$ represents the vacuum field
state), 
where $|\alpha\rangle$ is a coherent state of large amplitude: $|\alpha|\gg
1.$ We take $\alpha = i\,{\dot q}(0)\sqrt{M/2\hbar \omega_0}, $ so that $%
|\alpha\rangle$ is a `semiclassical' state associated to a minimum
uncertainty wave-packet whose initial velocity is $\dot q(0).$ 
We have shown in Sec.~IV that the action of the vacuum radiation pressure
on the motion of the mirror is a very small perturbation
(weak coupling limit). Thus
the time evolution may be computed from a simple `semi-classical' model, in which the
field evolution is obtained assuming the classical {\it prescribed} motion 
\begin{equation}
{\dot q}(t)={\dot q}(0) \cos(\omega_0 t),  \label{q}
\end{equation}
where $q(t)$ is the position of the mirror at time $t.$ The dynamical
Casimir effect is described by the interaction Hamiltonian (see Ref.~\cite{barton}, and 
compare with the first term in Eq.~(\ref{Hint})) 
\begin{equation}
H_{{\rm int}}= -{\dot q}(t) {\cal P}.
\end{equation}
The amplitude $b(\Delta\! t)$ 
for the creation of photon pairs corresponding to 
frequencies $\omega_1$ and $\omega_2$ at time $\Delta\! t$
is given by 
\begin{equation}
b(\omega_1,\omega_2;\Delta\! t) ={\frac{i}{\hbar}}\langle
\omega_1,\omega_2|{\cal P}|0\rangle
 \int_0^{\Delta t} dt^{\prime}e^{i(\omega_1+\omega_2)t^{\prime}} {%
\dot q}(t^{\prime}).  \label{b1}
\end{equation}
According to Eq.~(\ref{b1}), the amplitude depends on the {\it sign} of ${%
\dot q},$ which is very important to the discussion of decoherence. 

Replacing Eq.~(\ref{q}) into (\ref{b1}), we find for the two-photon
probabilities 
\begin{equation}
|b(\omega_1,\omega_2;\Delta\! t)|^2\approx {\frac{1}{\hbar^2}} |\langle
0|{\cal P}|\omega_1,\omega_2\rangle|^2 {\dot q}(0)^2  \label{b2}
\end{equation}
\[
\times{\frac{\sin^2\left[(\omega_1+\omega_2-\omega_0)\Delta\! t/2\right]}{%
(\omega_1+\omega_2-\omega_0)^2}}.
\]
For $\omega_0 \Delta\! t \gg 1,$ the r.-h.-s. of Eq.~(\ref{b2}) is sharply
peaked around $\omega_1+\omega_2=\omega_0.$ Thus, for large times energy is
well defined, in agreement with the time-energy uncertainty relation. In
this limit, Eq.~(\ref{b2}) yields 
\begin{equation}
|b(\omega_1,\omega_2;\Delta\! t)|^2\approx {\frac{\pi}{2\hbar^2}} |\langle
0|{\cal P}|\omega_1,\omega_2\rangle|^2 {\dot q}(0)^2\, \Delta\! t \label{b3}
\end{equation}
\[
\times\delta(\omega_1+\omega_2-\omega_0) \nonumber
\]
Since the source of the radiated energy is the motion of mirror, 
one may expect that the two-photon probabilities are related to the
amplitude decay rate $\Gamma.$ The radiated energy during $\Delta t$ is
\begin{equation}
\Delta\! E = {\frac{1}{2}} \int_0^{\infty}d\omega_1 \int_0^{\infty}d\omega_2
|b(\omega_1,\omega_2;\Delta\! t)|^2 \hbar (\omega_1+\omega_2),  \label{DE}
\end{equation}
which according to Eq.~(\ref{b3}) is proportional to the time interval $\Delta\! t.$
The energy of the mirror decays as 
$d E_M/d t  = - 2 \Gamma E_M,$ where $E_M=M {\dot q}(0)^2/2.$
Hence, from energy conservation we have 
$$\Gamma = {1\over M {\dot q}(0)^2} {\Delta\! E\over \Delta\! t},$$
leading, with the help of (\ref{b3}) and (\ref{DE}), to the representation given by~(\ref{Gammaa}).

To analyze the effect of decoherence, we take the field to be initially in the 
`even' superposition state 
$|\psi\rangle_{\rm e} = (|\alpha \rangle +|-\alpha \rangle )/\sqrt{2},$
so that the mirror-plus-field state at $t=0$ is 
\[
|\Psi\rangle_0 = |\psi\rangle_{\rm e} \otimes |0\rangle. 
\]
By linearity, its time evolution is obtained from the two--photon 
amplitudes given by~(\ref{b1}):
\begin{equation}
|\Psi\rangle_{\Delta\! t} = (|\alpha\rangle \otimes |\phi^{+}\rangle_{\Delta\! t} +|-\alpha\rangle
\otimes |\phi^{-}\rangle_{\Delta\! t} )/\sqrt{2}, \label{Psi1}
\end{equation}
where 
\begin{equation}
|\phi^{\pm}\rangle_{\Delta\! t}= B({\Delta\! t})|0\rangle \pm {\frac{1}{2}} \int_0^{\infty}d%
\omega_1\int_0^{\infty}d\omega_2\, b(\omega_1,\omega_2;{\Delta\! t})
|\omega_1,\omega_2\rangle.\label{phipm}
\end{equation}
The already noted sensitivity of the two-photon amplitudes to the phase of 
the motion of the mirror, which is explicit through the `minus' sign for
$|\phi^{-}\rangle$ in~(\ref{phipm}), generates entanglement 
between mirror and field. This is discussed in Sec.~V, whose starting point is
Eq.~(\ref{Psit}), which is derived by replacing (\ref{phipm}) into (\ref{Psi1}).



\begin{references}

\bibitem{physicstoday}  W. H. Zurek, Phys. Today {\bf 44}, No. 10, 36 (1991).

\bibitem{monroe} C. Monroe, D. M. Meekhof, B. E. King, and D. J. Wineland,
Science {\bf 272}, 1131 (1996). 

\bibitem{haroche}  M. Brune, E. Hagley, J. Dreyer, X. Ma\^{i}tre, A. Maali,
C. Wunderlich, J. M. Raymond, and S. Haroche, Phys. Rev. Lett. {\bf 77}, 4887 (1996).

\bibitem{luiz} L. Davidovich, M. Brune, J. M. Raimond, and S. Haroche,
Phys. Rev. A {\bf 53}, 1295 (1996).

\bibitem{wineland} C. J. Myatt, B. E. King, Q. A. Turchette, C. A. Sackett,
D. Kielpinski, W. M. Itano, C. Monroe, and D. J. Wineland, Nature {\bf 403},
269 (2000).

\bibitem{caldeira}  A. O. Caldeira and A. J. Leggett, Phys. Rev. {\bf A} 31,
1059 (1985).

\bibitem{all}  W. G. Unruh and W. H. Zurek, Phys. Rev. D {\bf 40}, 1071
(1989); B. L. Hu, J. P. Paz and Y. Zhang, Phys. Rev. D {\bf 45}, 2843 (1992).

\bibitem{review} M. Kardar and R. Golestanian, Rev. Modern Phys. {\bf 71}, 1233
 (1999), and references therein.

\bibitem{spectrum} P. A. Maia Neto and L. A. S. Machado, Phys. Rev. A 
{\bf 54}, 3420 (1996).

\bibitem{diego-Casimir}  D. A. R. Dalvit and F. D. Mazzitelli, Phys. Rev. A 
{\bf 59}, 3059 (1999).

\bibitem{ford}  L. H. Ford and A. Vilenkin, Phys. Rev. D {\bf 25}, 2569
(1982).

\bibitem{force-Casimir} M. T. Jaekel and S. Reynaud, 
Quantum Optics {\bf 4}, 39 (1992); 
G. Barton and C. Eberlein, Ann. Phys. (NY) {\bf 227}, 222 (1993);
P. A. Maia Neto, J. Phys. A: Math. Gen. {\bf 27}, 2167
(1994). 

\bibitem{prl}  D. A. R. Dalvit and P. A. Maia Neto, Phys. Rev. Lett. {\bf 84}, 
798 (2000).

\bibitem{jr}  M. T. Jaekel and S. Reynaud, J. Phys. France {\bf I 3}, 1
(1993).

\bibitem{jr2}  M. T. Jaekel and S. Reynaud, Phys. Lett. {\bf A 180}, 9
(1993).

\bibitem{barton}  G. Barton and A. Calogeracos, Ann. Phys. (NY) {\bf 238},
227 (1995).

\bibitem{barton2} A. Calogeracos and G. Barton, Ann. Phys. (NY) {\bf 238}, 268
(1995).

\bibitem{walls} D. F. Walls and G. Milburn, Quantum Optics, Springer
Verlag 1995.

\bibitem{zurek} W. H. Zurek, Prog. Theor. Phys. {\bf 89}, 281 (1993).

\bibitem{Paz}  W. H. Zurek, S. Habib and J. P. Paz, Phys. Rev. Lett. 
{\bf 70}, 1187 (1993).

\bibitem{wiseman} H. M. Wiseman and J. A. Vaccaro, Phys. Lett. A {\bf 250},
241 (1998).

\bibitem{foot1} The contribution associated to $D_2$ is smaller by a 
factor of the order of $1/|\alpha|.$

\bibitem{pamn93}  P. A. Maia Neto and S. Reynaud, Phys. Rev. A {\bf 47}.
1639 (1993).

\bibitem{barton91}  G. Barton, J. Phys. A Math. Gen. {\bf 24}, 5533 (1991).
 
\bibitem{kubo}  R. Kubo, Rep. Progr. Phys. {\bf 29}, 255 (1966).

\bibitem{JRT}  M. T. Jaekel and S. Reynaud, Phys. Lett. A {\bf 172}, 319
(1993).

\bibitem{energy}  J. P. Paz and W. H. Zurek, Phys. Rev. Lett. {\bf 82}, 5181
(1999).


\bibitem{runaway} It is known from classical electron theory that 
this differential
equation suffers from instabilities associated to `runaway' solutions.
However, the model for the frequency-dependent reflection amplitudes considered here satisfies 
the conditions of passivity associated to mechanical stability [M.-T. Jaekel and S. Reynaud, 
Phys. Lett. A {\bf 167}, 227 (1992)]. Although Eq.~(\ref{run}) is a 
useful shortcut for cross-checking 
the result for $\Gamma$ as given by Eq.~(\ref{Gammaper}),
 it is certainly not
equivalent to the more rigorous approach employed in Secs.~II and V, where 
the perfectly-reflecting limit is taken only at the last step. 

\bibitem{foot2} This discrepancy was already note in Ref.~\cite{JRT}. 
According to this reference, it originates from neglecting the effect of 
amplitude modification of the reflected thermal field due to the motion of the mirror.  

\bibitem{foot3} Jointly with the conditions $ \lambda_{\rm th}^2\ll A$ 
and $\lambda_{\rm th}\ll c/\omega_0$ (high temperature limit),
it imposes $\Delta\! Q^2 \ll A.$ Thus,  a necessary condition for
the validity of~(\ref{restemp}) is that
the distance between the wavepackets must
be smaller that the `size' of the mirror. 


\end{references}
\end{document}